\documentclass[10pt,oneside,english]{article}
\usepackage{geometry}
\usepackage{graphics}
\usepackage{setspace}
\usepackage{subfigure}
\newcommand {\car}{$^{12}$C }
\newcommand {\oxy}{$^{16}$O }
\newcommand {\caI}{$^{40}$Ca }
\newcommand {\caII}{$^{48}$Ca }
\newcommand {\lead}{$^{208}$Pb }
\newcommand{\br}{{\bf r}}

\newcommand{\bsigma}{\mbox{\boldmath $\sigma$}}
\newcommand{\btau}{\mbox{\boldmath $\tau$}}

\begin{document}
\title{\bf New results in the CBF theory\\
for medium-heavy nuclei}
\author{C. Bisconti$^{\,1,2}$, 
        G. Co'$^{\,1}$ and F. Arias de Saavedra$^{3}$\\
 1) Dipartimento di Fisica, Universit\`a di Lecce,\\
 and I.N.F.N. sezione di Lecce,\\
 I-73100 Lecce, Italy\\
 2) Dipartimento di Fisica, Universit\`a di Pisa, 
 I-56100 Pisa, Italy \\
 3) Departamento de F\'{\i}sica
 At\'omica, Molecular y Nuclear,\\
 Universidad de Granada, 
 E-18071 Granada, Spain }
\maketitle
\begin{abstract}
  Momentum distributions, spectroscopic factors and quasi-hole wave
  functions of medium-heavy doubly closed shell nuclei have been
  calculated in the framework of the Correlated Basis Function theory,
  by using the Fermi hypernetted chain resummation techniques.  The
  calculations have been done by using microscopic two-body
  nucleon-nucleon potentials of Argonne type, together with three-body
  interactions. Operator dependent correlations, up to the tensor
  channels, have been used. 
\end{abstract}
%
%

\section{Introduction}
\label{sec:introduction}
The validity of the non relativistic description of the atomic nuclei
has been well established in the last ten years. The idea is to describe
the nucleus with a Hamiltonian of the type:
\begin{equation}
H=-\frac{\hbar^2}{2m}\sum_{i}\nabla^2+ 
\sum_{i<j}v_{ij}+\sum_{i<j<k}v_{ijk}  \,\,\,,
\label{eq:hamiltonian}
\end{equation}
where the two- and three-body interactions, $v_{ij}$ and $v_{ijk}$
respectively, are fixed to reproduce the properties of the two- and
three-body nuclear systems. 

About fifteen years ago, we started a project aimed to apply to the
description of medium and heavy nuclei the Correlated Basis Function
(CBF) theory, successfully used to describe the nuclear and neutron
matter properties \cite{wir88,akm98}.  We solve the many-body
Schr\"odinger equation by using the variational principle:
\begin{equation}
\delta E[\Psi]=
\delta \frac{<\Psi|H|\Psi>}{<\Psi|\Psi>} = 0  \,\,\,.
\label{eq:varprin}
\end{equation}
The search for the minimum of the energy functional is carried out within a
subspace of the full Hilbert space spanned by the $A$-body wave
functions which can be expressed as:
\begin{equation}
\Psi(1,...,A)={\cal F}(1,...,A)\Phi(1,...,A)  \,\,\,,
\label{eq:psi}
\end{equation}
where ${\cal F}(1,...,A)$ is a many-body correlation operator, and
$\Phi(1,...,A)$ is a Slater determinant composed by single
particle wave functions, $\phi_{\alpha}(i)$. We use two and
three-body interactions of Argonne and Urbana type, and we consider
all the interaction channels up to the spin-orbit ones. The complexity
of the interaction requires an analogously complex correlation:
\begin{equation}
{\cal F}={\cal S}\left( \prod_{i<j=1}^{A}F_{ij} \right) \,\,\,,
\label{eq:cor1}
\end{equation}
where ${\cal S}$ is a symmetrizer operator and $F_{ij}$ has the form:
\begin{equation}
F_{ij}=\sum_{p=1,6}f^p(r_{ij})O^p_{ij}  \,\,\,.
\label{eq:cor2}
\end{equation}
In the above equation we have adopted  the nomenclature commonly used
in this field, by defining the operators as:
\begin{equation}
O^{p=1,6}_{ij}=[1,\bsigma_i\cdot\bsigma_j,S_{ij}]\otimes
[1,\btau_i\cdot\btau_j]  \,\,\,,
\end{equation}
where
 $S_{ij}=(3\hat{\br}_{ij}\cdot\bsigma_i\hat{\br}_{ij}\cdot\bsigma_j
-\bsigma_i\cdot\bsigma_j)$ is the tensor operator.

The binding energies and the charge distributions of \car, \oxy, \caI,
\caII and \lead doubly closed shell nuclei have been obtained by
solving the Fermi Hypernetted Chain (FHNC) equations in the Single
Operator Chain (SOC) approximation \cite{bis06}.  These calculations
have the same accuracy of the best variational calculations done in
nuclear and neutron matter \cite{wir88,akm98}.

By using this FHNC/SOC computational scheme we have investigated the
effects of the correlations on some ground state quantities which are
related to observables.  A first quantity we have studied is the
momentum distribution. It is well known that the short--range correlations
enhance by orders of magnitude the high-momentum components of the
momentum distribution \cite{ant88}.  We have also studied the
spectroscopic factors since, as we have already mentioned, their
empirical values are always smaller than the independent particle
model (IPM) predictions.

The present study has been done by using the many-body wave functions
obtained in Ref. \cite{bis06}. The Argonne $v'_{8}$ two-nucleon
potential, together with the Urbana XI three-body force has been used.

In the next sections we present the results of our study of the
One-Body Density Matrix (OBDM) and of the momentum distribution and 
we evaluate the spectroscopic factors.
\section{The momentum distributions}
The OBDM, of a system of $A$ nucleons is defined as:
\begin{eqnarray}
\nonumber
&~& 
\rho^{s,s';t}(\br_1,\br'_{1}) = \frac{A}{<\Psi|\Psi>} 
\int dx_2  \ldots dx_A  \\
&~&  
\Psi^{\dagger}(x_1,x_2,\ldots,x_A) \chi_t(1) \chi_s(1)
\chi_t^\dagger(1') \chi_{s'}^\dagger(1')\Psi(x'_1,x_2,\ldots,x_A) \,\,\,.
\label{eq:obdm}
\end{eqnarray}
In the above expression, the variable $x_i$ indicates the position
($\br_i$) and the third components of the spin $(s)$ 
and of the isospin $(t)$  
of the single nucleon.  With the integral sign we understand that also
the sum on spin and isospin of all the particles, including $1'$, is 
performed.  The OBDM of eq. (\ref{eq:obdm}) is characterized by the
quantum numbers relative to the particle $1$.  
In our calculations we are interested in the quantity:
\begin{equation}
\rho^t(\br_1,\br'_{1}) =
\sum_{s=\pm 1/2} \left[
\rho^{s,s;t}(\br_1,\br'_{1}) + \rho^{s,-s;t}(\br_1,\br'_{1})
\right]  \,\,\,,
\label{eq:obdm1}
\end{equation}
whose diagonal part represents the one-body density of neutrons
or protons, this last one is related to the charge density
distribution of the nucleus. 

We define 
the momentum distributions of protons or neutrons as: 
\begin{equation}
n^t(k)= \frac {1} {{\cal N}_t} \int d\br_1 d\br'_1 \, 
e^{i {\bf k} \cdot (\br_1-\br'_1)} 
\rho^t(\br_1,\br'_1) \,\,\,,
\label{eq:md}
\end{equation}
where we have indicated with ${\cal N}_t$  the number of protons 
or neutrons. 
We describe doubly closed shell nuclei, with different numbers of
proton and neutrons, in a $jj$ coupling scheme. 
\begin{figure}[pt]
{\centering \resizebox*{11.cm}{!}{\includegraphics{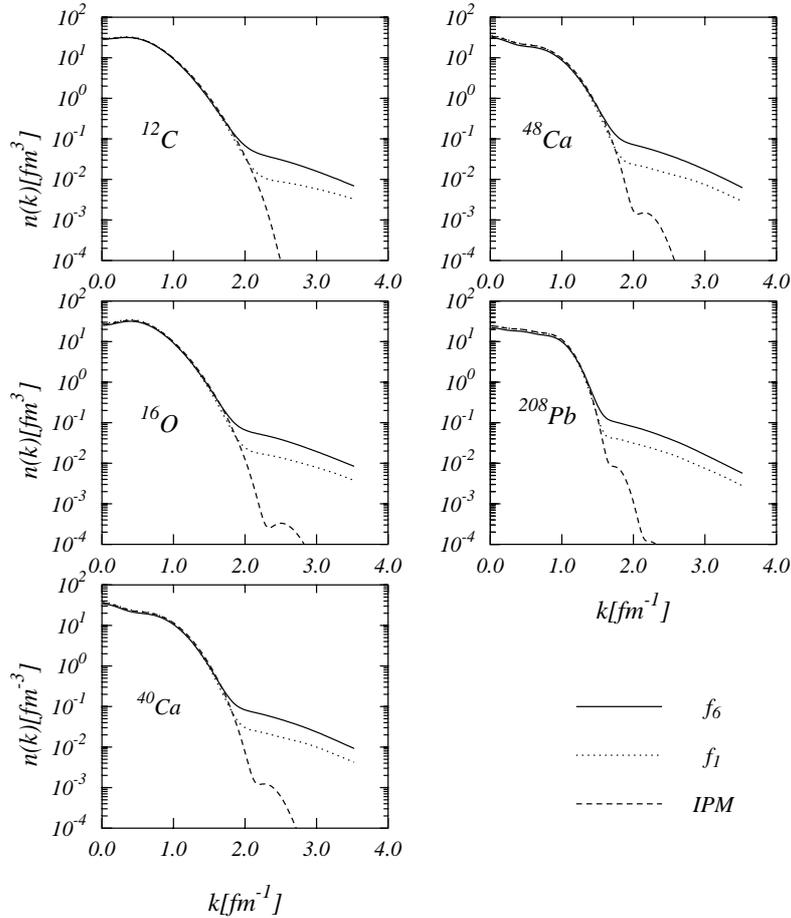}} \par}
\caption{\small The proton  momentum distributions 
for the \car, \oxy, \caI, \caII, \lead
nuclei calculated in the IPM model ( dashed lines), by using the
scalar correlation only (dotted lines) and the full correlation
operator in FHNC/SOC approximation (solid lines).}
\label{fig:md}
\end{figure}
The correlated OBDM is obtained by using the ansatz (\ref{eq:psi}) in
Eq. (\ref{eq:obdm}). 
The calculation is done by following the lines indicated in
Ref. \cite{fab01} and considering, in addition, the presence of the
antiparallel spin terms and distinguishing proton and neutron
contributions.  The diagonal part of the OBDM is the one-body density,
normalized to the number of nucleons.  We present in Fig.\ref{fig:md}
the proton momentum distributions, for the \car, \oxy, \caI, \caII and
\lead nuclei. The dashed lines show the results obtained in the IPM
model, the dotted lines those results calculated by using the scalar
correlations only and, finally, the solid lines indicate the momentum
distributions obtained by using the full operator dependent
correlations.  The high momentum components of the correlated
distributions are orders of magnitude larger than those produced by
the IPM. The operator dependent terms of the correlations further
increase this behavior.
\section{The spectroscopic factors}
\label{sec:specf}
The quasi-hole wave function is defined as:
\begin{equation}
\psi_{nljm}^{t}(x)=
\sqrt{A}\frac{<\Psi_{nljm}(A-1)|\delta(x-x_A)P^{t}_{A}|\Psi(A)>}
{<\Psi_{nljm}(A-1)|\Psi_{nljm}(A-1)>^{1/2}
<\Psi(A)|\Psi(A)>^{1/2}} \,\,\,,
\label{eq:qhfun}
\end{equation}
where we have indicated with $\Psi_{nljm}(A-1)$ and $\Psi(A)$ the wave
functions describing the nuclei formed by $A-1$ and $A$ nucleons
respectively, and with $P^t_A$ the isospin projector. The subindexes
$nljm$ designate the quantum numbers of the odd-even nucleus.

We describe the wave function of the nucleus with $A-1$ nucleons by
using an ansatz analogous to that of Eq. (\ref{eq:psi}):
\begin{equation}
\Psi_{nljm}(A-1)=\, F(1,...,A-1) \, \Phi_{nljm}(1,...,A-1)
\,\,\,,
\label{eq:psiam1}
\end{equation}
\begin{figure}[pt]
{\centering \resizebox*{10cm}{!}{\includegraphics{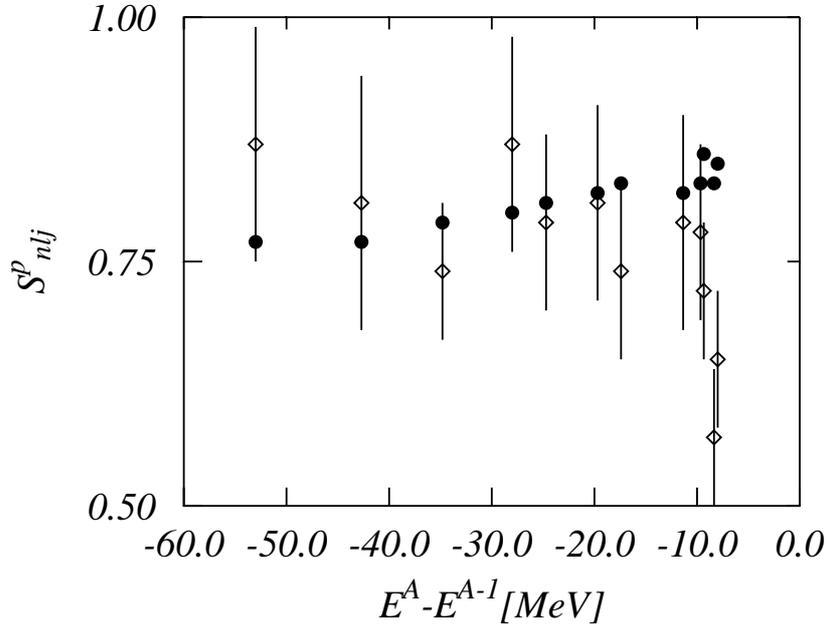}} \par}
\caption{\small Comparison between the theoretical,
black points, and the experimental, empty diamonds,  
spectroscopic factors for some proton states of  
$^{208}Pb$. In the $x$ axis we indicate the
separation energies.}
\label{fig:sf} 
\end{figure}
where $\Phi_{nljm}(1,...,A-1)$ indicates a Slater determinant obtained
by removing from the Slater determinant $\Phi(1,...,A)$ a single state
characterized by the quantum numbers $nljm$.  In the IPM 
the quasi-hole wave functions coincide with the mean-field wave
functions.

In order to obtain the radial part of the quasi-hole wave function we
multiply equation (\ref{eq:qhfun}) by
$< jm |$, we integrate over the angular
coordinate $\Omega$, and sum on the spin coordinates \cite{fab01}:
\begin{equation}
\psi_{nlj}^{t}(r)= \sum_{\mu,s} <l\mu 1/2 s |jm>
\int d\Omega \,
Y_{l\mu}^*(\Omega)\chi^{\dagger}_{s} \, \psi_{nljm}^{t}(x)
= {\cal X}_{nlj}^t(r)[{\cal N}_{nlj}^t]^{1/2}
\,\,\,.
\label{eq:qhprod}
\end{equation}

From the knowledge of the quasi-hole functions we obtain the
spectroscopic factors as:
\begin{equation}
S_{nlj}^t=\int dr\,r_1^2\,|\psi_{nlj}^t(r)|^2
\,\,\,.
\label{eq:sf}
\end{equation}

In Fig. \ref{fig:sf} we compare the theoretical spectroscopic factors
calculated for the proton bound states of the \lead nucleus with the
experimental data of Ref. \cite{van01t}.  In abscissa we give the
separation energies defined as the difference between the energy of a
$A$-nucleon system and that of the $A-1$-nucleon system obtained by
removing the $nljm$ state.  The agreement between theory and experiment
is better for the deeply bound shells than for those levels closer to
the Fermi surface. This could be due to the strong coupling between
the quasi-hole wave function and the low-lying surface vibrations.
The effects of this coupling, usually called long-range correlations,
not explicitly treated by our theory, are expected to be larger for
the external shells than for the internal ones.

\begin{figure}[pt]
{\centering \resizebox*{14cm}{!}{\includegraphics{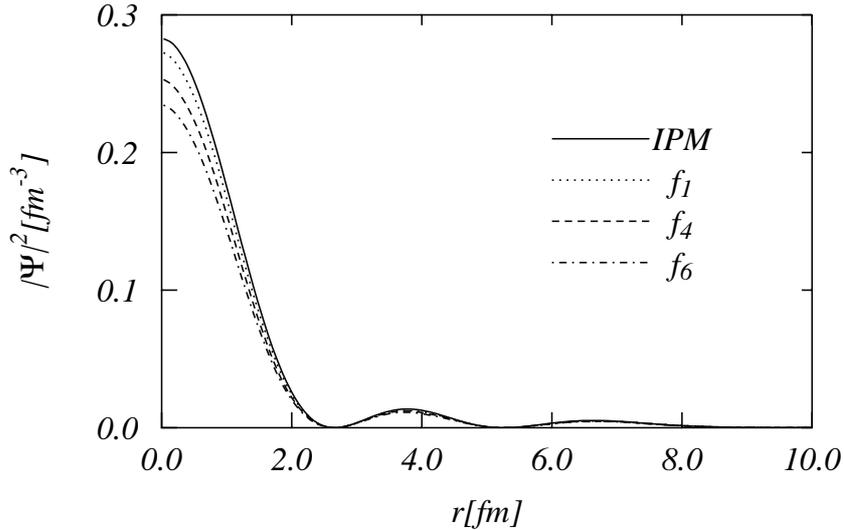}} \par}
\caption{The square of the 
quasi-hole wave function for the $3s_{1/2}$
proton state of the \lead. The labels $unc$, $Jas$, $f^4$ and $f^6$
indicate the IPM,Jastrow, $f^4$ and $f^6$ models respectively.}
\label{fig:qh_fun}
\end{figure}

The effect of the correlations on the quasi-hole wave functions is
presented in Fig. \ref{fig:qh_fun} where we have shown the squared
quasi-hole $3s_{1/2}$ proton wave function calculated with increasing
complexity in the correlation function.  The full line indicates the
IPM result, the other lines have been obtained by using only scalar
correlations, $f_1$, operatorial correlations without the tensor
channels, $f_4$, and correlations which include also the tensor
dependent terms, $f_6$.  The presence of the correlations produces a
lowering of the quasi-hole wave function in the nuclear center. There
is a consistent trend of the correlations effects: the more elaborated
is the correlations the larger is the decreasing at the center of the
nucleus.
\section{Summary and conclusions}
\label{sec:summary}
In this work momentum distributions, spectroscopic factors and
quasi-hole wave functions of medium-heavy doubly closed shell nuclei
have been calculated by extending the FHNC/SOC computational scheme.
The calculations have been done considering the different number of
proton and neutrons and the single particle basis 
are given in a $jj$ coupling scheme. A microscopic two-body
interaction of Argonne type, implemented with the appropriated
three-body force of Urbana type, have been used.
The calculations have been done with operator dependent correlations
which include, in addition to the four central channels, also tensor
correlations.  

The comparison between our results with those obtained in the IPM
highlights the correlations effects. The correlated momentum
distributions have high momentum tails which are orders of magnitude
larger than the IPM results. The spectroscopic factors are always
smaller than one, the IPM value, and in a reasonable agreement with
the experimental values, especially for the more bounded states.  The
quasi-hole wave functions are depleted in the nuclear center by the
correlations. We have found that the operator dependent terms emphasize the
correlations effects.
\vskip 0.5 cm
\begin{center}
{\bf Aknowledgments}
\end{center}
This work has been partially supported by the agreement INFN-CICYT, by
the Spanish Ministerio de Educaci\'on y Ciencia (FIS2005-02145) 
and by the MURST through the PRIN: {\sl Teoria della struttura dei nuclei 
e della materia nucleare}.
\bibliographystyle{ws-procs9x6}

\begin{thebibliography}{1}
  
\bibitem{wir88} R.~B. Wiringa, V.~Ficks and A.~Fabrocini, {\em Phys.\ 
    Rev. \ C} {\bf 38}, p.  1010 (1988).

\bibitem{akm98} A.~Akmal, V.~Pandharipande and D.~G. Ravenhall, {\em
    Phys. \ Rev. \ C} {\bf 58}, p. 1804 (1998).

\bibitem{bis06} C.~Bisconti, F.~Arias de~Saavedra, G.~Co' and
  A.~Fabrocini, {\em Phys. \ Rev. \ C} {\bf 73}, p. 054304 (2006).
  
\bibitem{ant88} A.~N. Antonov, P.~E. Hodgson and I.~Z. Petkov, {\em
    Nucleon momentum and density distributions} (Clarendon, Oxford,
  1988).
  
\bibitem{fab01} A.~Fabrocini and G.~Co', {\em Phys.\ Rev. \ C} {\bf
    63}, p. 044319 (2001).

\bibitem{van01t} M.~F. van Batenburg, 
  {\em Deeply bound protons in $^{208}$Pb},
  PhD thesis, Universiteit Utrecht (Nederlands) 2001,
  unpublished.

\end{thebibliography}
%

\end{document}